\documentclass{sf2a-conf}
\usepackage{graphicx}

\begin{document}
\TitreGlobal{SF2A 2008}
\title{The WFI H$\alpha$ spectroscopic survey}
\author{Martayan, C.$^{1,}$} \address{Royal Observatory of Belgium, 3 avenue circulaire 1180 Brussels, Belgium}
\address{GEPI, Observatoire de Paris, CNRS, Universit\'e Paris Diderot; 5 place Jules Janssen 92195 Meudon Cedex, France}
\author{Baade, D.} \address{European Organisation for Astronomical Research in the Southern Hemisphere, Karl-Schwarzschild-Str. 2, D-85748 Garching b. Muenchen, Germany}
\author{Fabregat, J.} \address{Observatorio Astron\'omico de Valencia, edifici Instituts d'investigaci\'o, Poligon la Coma, 46980 Paterna Valencia, Spain}
\runningtitle{The WFI H$\alpha$ spectroscopic survey}
\setcounter{page}{237} 

\index{Martayan, C.}
\index{Baade, D.}
\index{Fabregat, J.}

\maketitle
\begin{abstract}
This document presents the results from our spectroscopic survey of H$\alpha$ emitters in galactic and SMC open
clusters with the ESO Wide Field Imager in its slitless spectroscopic mode. First of all, for the galactic
open cluster NGC6611, in which, the number and the nature of emission line stars is still the object of
debates, we show that the number of true circumstellar emission line stars is small. Second, at low
metallicity, typically in the Small Magellanic Cloud, B-type stars rotate faster than in the Milky Way and
thus it is expected a larger number of Be stars. However, till now, search for Be stars was only performed
in a very small number of open clusters in the Magellanic Clouds. Using the ESO/WFI in its slitless
spectroscopic mode, we performed a H$\alpha$ survey of the Small Magellanic Cloud. 3 million low-resolution spectra
centered on H$\alpha$ were obtained in the whole SMC. Here, we present the method to exploit the data and first
results for 84 open clusters in the SMC about the ratios of Be stars to B stars. 
\end{abstract}
%
\section{Observations, data-reduction, and spectroscopic analysis}
Observations were performed in 2002, at the 2.2m of the ESO at la Silla equiped with the Wide Field Imager in its
slitless  spectroscopic mode (Baade et al. 1999).
This kind of instrumentation is not sensitive to the ambient diffuse nebula and displays only emission lines, which
come from circumstellar matter like in the case of Be stars. Be stars are very fast rotating stars, which are
surrounded by a  circumstellar decretion disk.
This instrumentation allowed Martayan et al. (2008a) to find true cirumstellar emission line stars in the  Eagle
Nebula and NGC6611 open cluster located in the Milky Way, while slit-spectroscopic observations show strong nebular
lines. Only a small number of true emission line stars (less than 10) was found.

In the Milky Way, we used broad bandpass filter centered in H$\alpha$, but in the Magellanic Clouds due to the crowding 
of the fields, we used a narrow bandpass filter also centered in H$\alpha$.
The exposure times range from 120 to 600s, and the resolving power is low ($\sim$100). 
In the SMC, 14 images were obtained, $\sim$8000 spectra were treated in 84 SMC open clusters 
among the 3 million obtained for the whole SMC, and in NGC6611 $\sim$10000 spectra were treated. 
In the LMC, 5 million spectra were obtained.   

\begin{figure}[ht]
\begin{center}
        \resizebox{8cm}{!}{\includegraphics[angle=-90]  {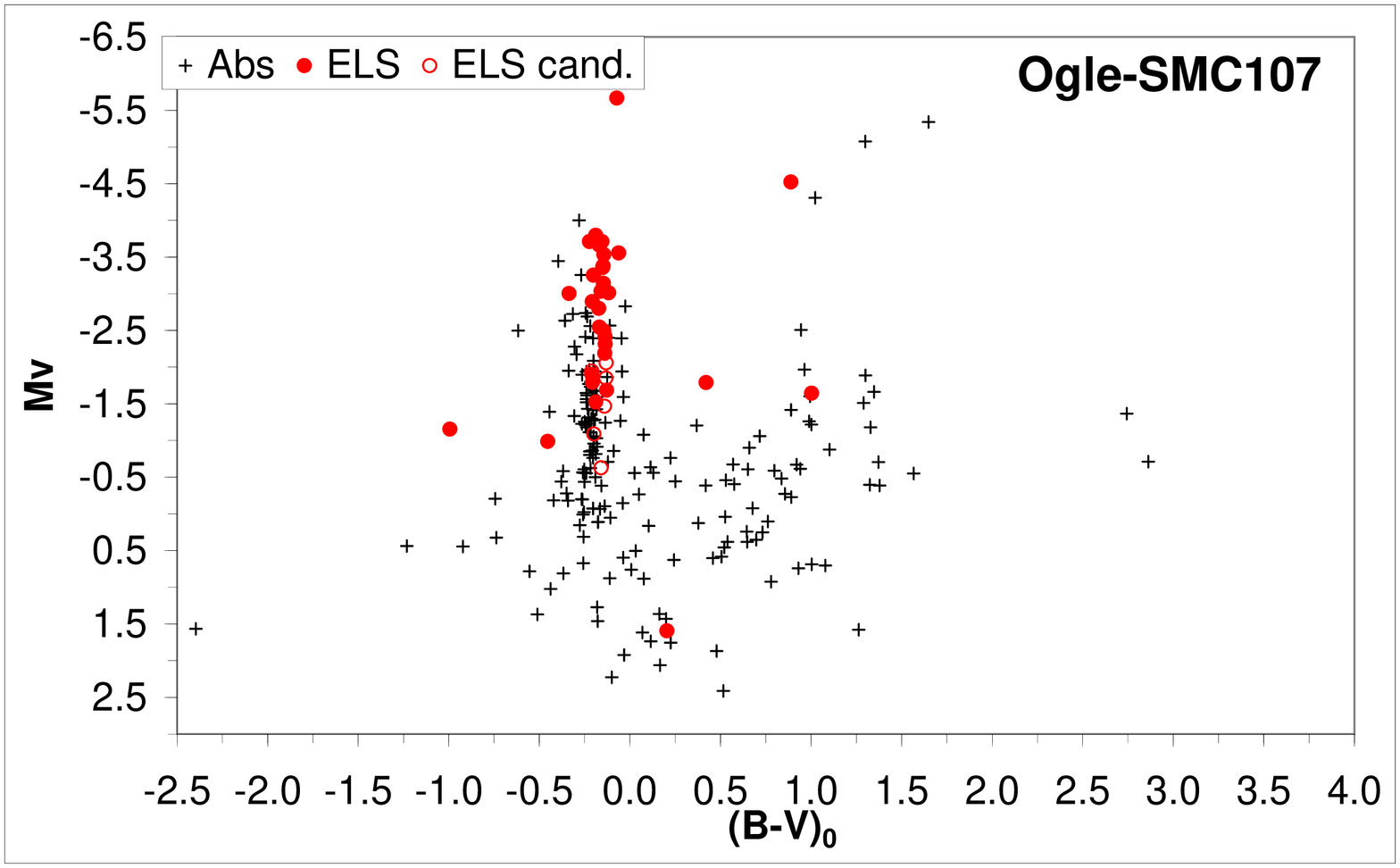}}
        \resizebox{8cm}{!}{\includegraphics[angle=-90]  {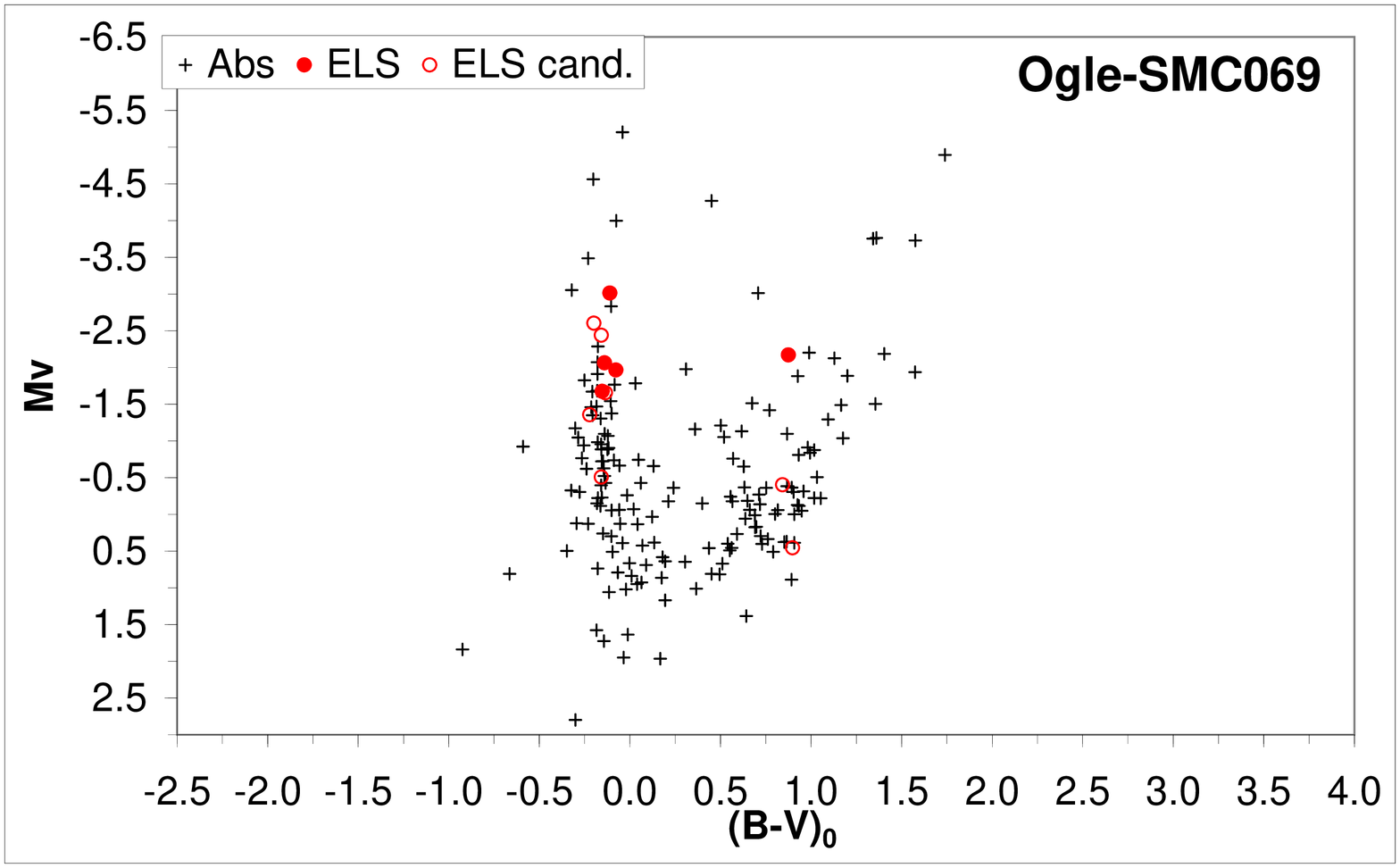}}
\caption{{\bf Left}: SMC open cluster OGLE-SMC107 (NGC330).  {\bf Right}: SMC open cluster OGLE-SMC069. 
Black dots correspond to the absorption stars, red full circles to definite emission line stars, and red empty circles to candidate
emission line stars.}
\label{fig1}
\end{center}
\end{figure}

The data-reduction was performed using IRAF tasks and the spectra extraction with SExtractor (Bertin \& Arnouts 1996).
The analysis of spectra and emission line stars detection were done using lecspec and ALBUM codes by Martayan et al. (2008a,b).
To classify the stars in SMC open clusters, we cross-correlated our WFI catalogues with OGLE ones (Udalski et al. 1998)
to obtain the photometry (B, V, I) for each star and various information for each open cluster (E[B-V], age, reddening).
More than 4000 stars of SMC open clusters were classified.
An example of colour-magnitude digrams is shown for different open clusters in the SMC in Fig.~\ref{fig1}.

\section{Metallicity effect}
At low metallicity, the stellar winds are less efficient than at high metallicity, as a consequence the stars retain more angular momentum
and rotate faster in the SMC than in the MW (Martayan et al. 2007, Hunter et al. 2008). It is then expected that there are more
Be stars in the SMC than in the MW.
To enlighten this potential effect, the stars from 83 of the 84 open clusters treated in the SMC were grouped in order to avoid the
variability of the rates of Be stars to B stars from a cluster to another. The rates of Be stars to B stars by spectral-type categories are
then compared with those obtained in the Milky Way with data from McSwain \& Gies (2005, 55 open clusters).
In each spectral-type category, the proportion of Be stars to B stars is twice to 4 times higher in the SMC than in the MW.
This result quantifies the trend seen in the preliminary studies of Maeder et al. (1999) or Wisniewski \& Bjorkman (2006).
About Oe stars, the rate is $\sim$1.5 times higher in the SMC than in the MW.

Furthermore, the distribution of the Be stars by spectral types peaks at the spectral-type B2 in the SMC open clusters.
The same behaviour is seen for early-type Be stars in the MW (Zorec \& Fr\'emat 2005 in the field or in open clusters 
with data from McSwain \& Gies 2005).

\section{Conclusion}
We conducted a large slitless spectroscopic survey in the Magellanic Clouds and in 2 open clusters in the Milky Way with the ESO/WFI in its
slitless spectroscopic mode. In the open cluster NGC6611 and the Eagle Nebula (M16), we show that there is only a small number of true
emission line stars. With the stars from 83 open clusters in the SMC, we show that there are twice to four times more Be stars in the SMC
than in the MW open clusters.
The exploitation of the spectra in the SMC field and in the whole LMC (field and open clusters) is currently ongoing.


\begin{acknowledgements}
C.M. acknowledges funding from the ESA/Belgian Federal Science Policy in the 
framework of the PRODEX program (C90290).
C.M. thanks support from ESO's DGDF 2006, 
\end{acknowledgements}

\end{document}